\documentclass[submitting]{nst}

\usepackage{subfigure,dcolumn}
\usepackage[T2A,T1]{fontenc}
\usepackage[russian,english]{babel}

\usepackage{listings}
\begin{document}

\title{Studies of Directed Flow with Event Plane Method in the HIRFL-CSR External-target Experiment}
\thanks{This work is supported in part by the National Key R\&D Program of China (No. 2024YFA1610700) and the National Natural Science Foundation of China (No. 12475147). }
\author{Wanlong Wu}
\affiliation{Department of Physics, Harbin Institute Of Technology, Harbin 150001, China}
\affiliation{Institute of Modern Physics, Chinese Academy of Sciences, Lanzhou 730000, China}
\author{Xionghong He}
\affiliation{Institute of Modern Physics, Chinese Academy of Sciences, Lanzhou 730000, China}
\author{Yanyu Ren}
\affiliation{Department of Physics, Harbin Institute Of Technology, Harbin 150001, China}
\author{Diyu Shen}
\affiliation{Institute of Modern Physics, Chinese Academy of Sciences, Lanzhou 730000, China}
\author{Shusu Shi}
\affiliation{Key Laboratory of Quark \& Lepton Physics (MOE) and Institute of Particle Physics,Central China Normal University, Wuhan 430079, China}
\author{Xu Sun}
\email[Corresponding author,]{xusun@impcas.ac.cn}
\affiliation{Institute of Modern Physics, Chinese Academy of Sciences, Lanzhou 730000, China}

\begin{abstract}

The Cooling-Storage-Ring External-target Experiment (CEE) at Heavy Ion Research Facility in Lanzhou (HIRFL) is designed to study the properties of nuclear matter created in heavy-ion collisions at a few hundred MeV/$u$ to 1 GeV/$u$ beam energies, facilitating the research of quantum chromodynamics phase structure in the high-baryon-density region.
Collective flow is one of the most important observables in heavy-ion collision experiments to study the bulk behavior of the created matter. 
Even though the standard event plane method has been widely used for collective flow measurements, it remains crucial to validate and optimize this method for the CEE spectrometer.
In this paper, we study  the experimental procedures of measuring directed flow in $^{238}$U+$^{238}$U collisions at 500  MeV/$u$ using event planes reconstructed by Multi Wire Drift Chamber and Zero Degree Calorimeter, respectively.  Jet AA Microscopic (JAM) transport generator is used to generate events, and the detector response is simulated by the CEE Fast Simulation (CFS) package.
Finally, the optimal kinetic region for proton directed flow measurements is discussed for the future CEE experiment.

\end{abstract}


\maketitle
\nolinenumbers

\section{Introduction}
\label{sec:intro}

Quantum chromodynamics (QCD) predicts a transition from hadronic matter to deconfined quark–gluon matter at sufficiently high temperature and/or high density~\cite{Shuryak:1978ij}. 
Heavy-ion collision experiments at Relativistic Heavy Ion Collider (RHIC) and the Large Hadron Collider (LHC) have provided unique experimental evidence for this transition~\cite{BRAHMS:2004adc,PHOBOS:2004zne,STAR:2005gfr,PHENIX:2004vcz}.
While striking progress has been made in the past decades, some foundational questions remain to be determined, such as the existence of critical end point in QCD phase diagram and the equation of state of nuclear matter at baryon densities much larger than the saturation density~\cite{Luo:2017faz}. 
At HIRFL-CSR energies, a hadronic gas with densities reaching 2–3 times nuclear saturation density and temperatures around 40 MeV can be produced.
Experiments at these energies is vital to elucidate the properties of QCD in the low-temperature and high-baryon-density region~\cite{Guo:2024zij}.

The CEE spectrometer is designed to measure charged final-state particles in the fixed target heavy-ion collisions experiment at HIRFL-CSR~\cite{Hu:2023niz}.
It is the first comprehensive nuclear physics experimental research platform in the GeV energy range in China\cite{Lu:2016htm}.
With the variety types of ion beam provided by HIRFL-CSR, e.g. $^{12}$C + $^{12}$C at 1.1 GeV/$u$ ($\sqrt{s_{\text{NN}}} = 2.36$ GeV) and $^{238}$U + $^{238}$U at 500 MeV/$u$ ($\sqrt{s_{\text{NN}}} = 2.1$ GeV)~\cite{Xia:2002xpu}, CEE spectrometer is an ideal platform to explore the QCD phase diagram at high baron density region, to study the nuclear matter equation of state and to search for the existence of the critical end point.

Collective flow is one of the most important observables in relativistic heavy-ion collision experiments to study the bulk behavior of the created matter~\cite{Voloshin:2008dg}. The azimuthal anisotropy of emitted particles in the momentum space can be expanded in a Fourier series~\cite{Voloshin:1994mz}:
\begin{equation}
E \frac{d^{3} N}{d p^{3}}=\frac{1}{2 \pi} \frac{d^{2} N}{p_{\mathrm{T}} d p_{\mathrm{T}} d y}\left(1+\sum_{n=1}^{\infty} 2 v_{n} \cos \left(n\left(\phi-\Psi_{\text{RP}}\right)\right)\right),
\label{eq:flow}
\end{equation}
where $\Psi_{\text{RP}}$ is the azimuthal angle of reaction plane defined by the beam direction and impact parameter. The Fourier coefficients $v_{n}=\left\langle\cos[n(\phi_{i}-\Psi_{\text{RP}})]\right\rangle$ is the $n^{\rm th}$-order flow coefficient, and the bracket means the average over all particles and events. $v_{1}$ is also referred as directed flow which is sensitive to the compressibility of the dense matter created in the heavy ion collisions~\cite{Voloshin:2008dg}.

The reaction plane angle is not directly measurable in the heavy ion collision experiment, but one could use the observed event plane angle $\Psi_{\text{EP}}$ from the anisotropic flow itself as an estimation on an event-by-event basis. This is the so-called \textit{standard event plane method}~\cite{Poskanzer:1998yz}. The event plane method has been widely used in the collective flow analysis in the past decades~\cite{STAR:2024ujm,NA49:2003njx,PHENIX:2011yyh,STAR:2008ftz,ALICE:2012vgf,Racz:2024eyi,Wei:2024mgr} , but it is still crucial to optimize the method for the CEE experiment and to validate that the collective flow signal obtained from CEE spectrometer is reliable.

This paper is organized as follows. Section~\ref{sec:ceeFastSim} describes the setup of CEE detector and the simulation tools. Section~\ref{sec:epMethod} presents the reconstruction and correction method of event plane from different sub-detectors of CEE experiment. Section~\ref{sec:tpcFlow} discusses the results of simulated proton $v_{1}$ from $^{238}\text{U}$ + $^{238}\text{U}$ collisions at $\sqrt{s_{\text{NN}}} = 2.1$ GeV. A summary is presented in Sec.~\ref{sec:summary}.

\section{CEE Detector and Fast Simulation }
\label{sec:ceeFastSim}
\begin{figure}[!htb]
\includegraphics[width=0.9\columnwidth]{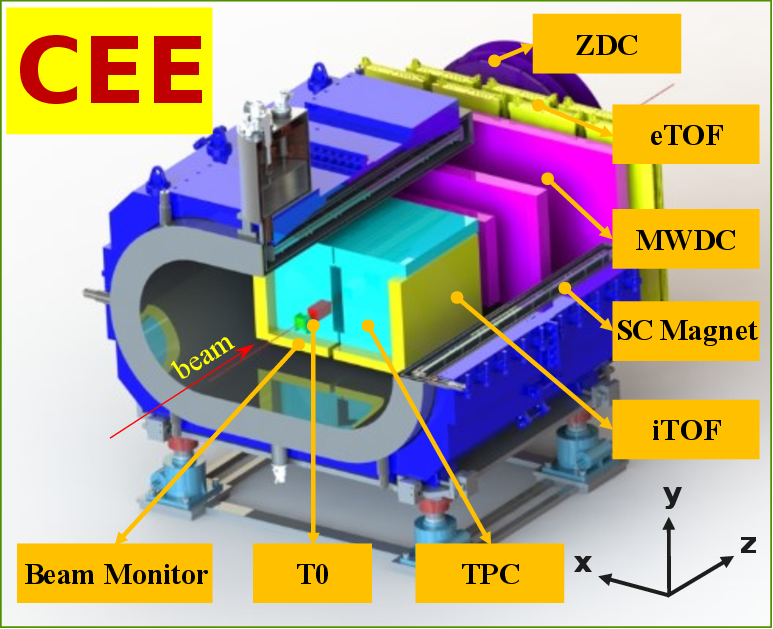}
\caption{The sketch of CEE detector\cite{Hu:2023niz}. }
\label{fig:ceesketch}
\end{figure}

The CEE spectrometer is an universal detection system designed for charged particle measurement of heavy-ion collisions in HIRFL-CSR energy region~\cite{Guo:2024zij}. Fig.~\ref{fig:ceesketch} presents the main detector configurations. The main components of CEE spectrometer are following: a large-gap dipole magnet with a 0.5 T magnetic field along the y-axis~\cite{Guo:2024zij}; a Time Projection Chamber (TPC)~\cite{article} system with two identical TPCs surrounded by the inner Time-of-Flight (iToF) ~\cite{Wang_2022} detectors locate in the mid-rapidity region; three layers of Multi Wire Drift Chamber (MWDC)~\cite{He:2024cew} followed by the end cap Time-of-Flight (eToF)~\cite{Wang_2020} detectors and Zero Degree Calorimeter (ZDC)~\cite{Zhu_2021} in the forward region; The start time (T0)~\cite{Hu_2017,Hu:2019mgr} detector and a silicon pixel positioning detector (BM)~\cite{Wang:2022evq} to monitor the beam position are installed on the beam line in the upstream side of the target. 
In this paper, the JAM model~\cite{PhysRevC.102.024913} was used to generate simulated events of $^{238}\text{U}$ + $^{238}\text{U}$ collisions at $\sqrt{s_{\text{NN}}} = 2.1$ GeV followed by CFS package to simulate the CEE experiment setup.

In JAM model, the initial position of nucleon is sampled according to the distribution of nuclear density. All hadronic states, including resonances, are propagated in space-time with explicit trajectories. The inelastic hadron-hadron collisions are described using two approaches: at low energies, resonance production dominates, while at high energies, the color string picture becomes the primary mechanism.
There are two modes in the model: cascade and mean-field. In the cascade mode, hadrons and their excited states have straight trajectory in two-body collisions. 
The nuclear mean-field mode incorporates the interactions of hadrons with the nuclear medium and the equation of state, 
which is implemented based on the relativistic quantum molecular dynamics approach(RQMD)~\cite{Sorge:1995dp, Liu:2019nii}.

The final-state particles generated by the JAM model are processed through the CEE Fast Simulation (CFS) framework. This framework simulates the CEE detector environment and generates responses for all CEE sub-detectors. The CFS enhances computational efficiency through parametric modeling and analytically derived formulations, which collectively simulate critical sub-detector characteristics such as detector acceptance, momentum resolution, energy deposition, and particle flight time.  Each sub-system's resolution effects are implemented via Gaussian smearing of the true input values. This methodology systematically accounts for measurement uncertainties while maintaining an optimal computational efficiency.

\begin{figure*}[htb!]
\includegraphics[width=0.8\textwidth]{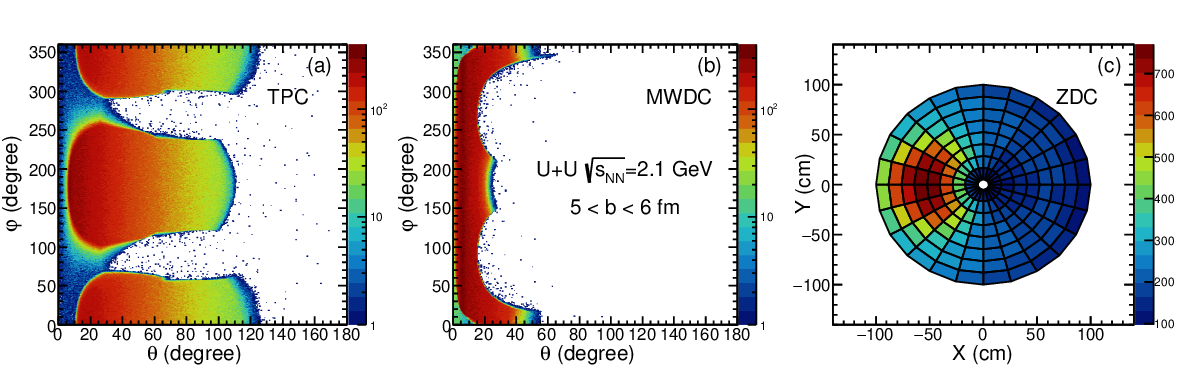}
\includegraphics[width=0.8\textwidth]{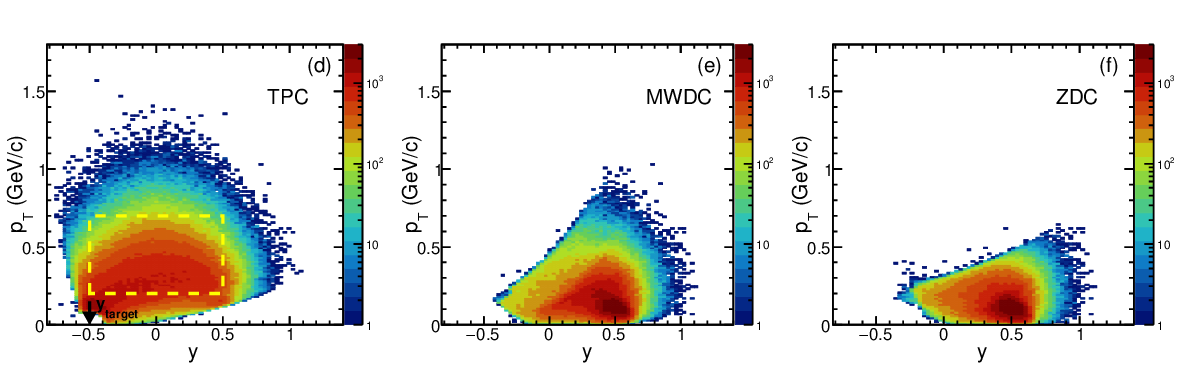}
\caption{Simulated proton acceptance from CFS package for 2.1 GeV $^{238}$U + $^{238}$U collisions with $5 < b < 6$ fm. Top panel: proton track/hit distribution in TPC (a), MWDC (b) and ZDC (c); Bottom panel: proton kinematic acceptance in TPC (d), MWDC (e) and ZDC (f). The dashed rectangle in panel (d) indicates the kinematic region used for the proton $v_{1}$ analysis.}
\label{fig:acpt}
\end{figure*}

Figure.~\ref{fig:acpt} shows the proton acceptance of TPC, MWDC and ZDC with events of impact parameter $5 < b < 6$ fm from JAM $+$ CFS simulation. 
The top panel of Fig.~\ref{fig:acpt} shows the angular coverage of TPC (Fig.~\ref{fig:acpt}(a))and MWDC (Fig.~\ref{fig:acpt}(b)) and the spacial coverage of ZDC (Fig.~\ref{fig:acpt}(c)).
CEE spectrometer roughly covers the polar angle from \SI{10}{\degree} to \SI{120}{\degree} in laboratory fame, corresponding to the proton rapidity range of -0.7 to 1 in the center-of-mass frame. Please note, all the rapidity ranges discussed in the following sections are in the center-of-mass frame. 
A clear efficiency lose in TPC azimuth at \SI{90}{\degree} and \SI{270}{\degree} is shown in Fig.~\ref{fig:acpt}(a), this is due to the two-half design of TPC~\cite{Huang:2018dus}. 
Fig.~\ref{fig:acpt}(c) shows the two dimensional X-Y hit distribution on ZDC. Clearly, left side ($X < 0$) of ZDC has much more hits than the right side ($X \geq 0$). The reason is that the final state charged particles are deflected by the magnetic field (along the y-axis), therefore, more likely to hit on one side of the ZDC~\cite{Liu:2023xhc}.
The bottom panel of Fig.~\ref{fig:acpt} shows the kinematic coverage of TPC, MWDC and ZDC. The dashed box in Fig.~\ref{fig:acpt}(d) indicates kinematic range ($0.2 < p_{T} < 0.7$ GeV/$c$ and $-0.5 < y < 0.5$) used in the directed flow simulation of charged particles measured by TPC. More details will be discussed in Sec.~\ref{sec:tpcFlow}.

\section{Event Plane Reconstruction with MWDC and ZDC}
\label{sec:epMethod}

The reaction plane in heavy-ion collisions is defined by the beam direction and impact parameter, which is not directly measurable. Experimentally, one could use the event plane method~\cite{Poskanzer:1998yz} to estimate the reaction plane, which uses emission azimuthal angle of detected particles to determine the event plane. 
The $n^{\rm th}$-order event plane angle, $\Psi_{n}$, could be calculated by the $n^{\rm th}$-order flow vector $\boldsymbol{Q}_{n}$. 
In this study, we focus on the $v_{1}$ simulation since it is more significant than higher order flow coefficient at CEE energy, thus, the $1^{\rm st}$-order event plane is used in the whole simulation process. The $1^{\rm st}$-order flow vector $\boldsymbol{Q}_{1}$ and event plane angle $\Psi_{1}$ are defiend as

\begin{equation}
\label{eq:qVec}
\begin{gathered}
	\boldsymbol{Q}_{1}=\left(\begin{array}{c} Q_{x,\text{1}}\\Q_{y,\text{1}} \end{array} \right)
	=\left( \begin{array}{c} \sum_{i}\omega_{i}\cos(\phi_{i})\\\sum_{i}\omega_{i}\sin(\phi_{i})\end{array} \right),\\
    \Psi_{1}=\tan^{-1}\frac{Q_{y,\text{1}}}{Q_{x,\text{1}}},
\end{gathered}
\end{equation}
where the sum goes over all particles used in the event plane determination and $\omega_{i}$ are weights to optimize the event plane resolution.

The main detectors used for event plane reconstruction in the CEE experiment are MWDC and ZDC. Since the MWDC is a track-based detector and ZDC is a hit-based detector, the information used to obtain the $\boldsymbol{Q}_{1}$ and the correction procedure are different. 

\begin{figure}[htbp]
\centering
\includegraphics[width=1.0\columnwidth]{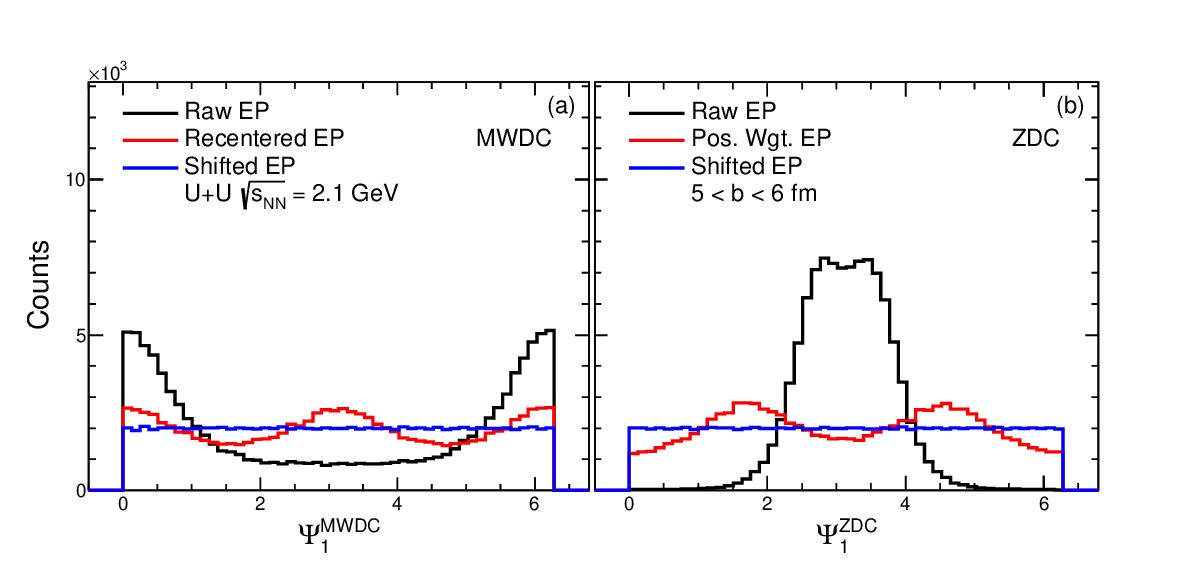}
\caption{(a) 1$^{st}$-order event plane distribution reconstructed by MWDC for raw distribution (black line), after re-center correction (red line) and after re-center $+$ shift correction (blue line); (b) 1$^{st}$-order event plane distribution reconstructed by ZDC for raw distribution (black line), after position weight correction (red line) and after position weight $+$ shift correction (blue line).}
\label{fig:ep}
\end{figure}

For track-based MWDC, the $\phi_{i}$ used in Eq.~\ref{eq:qVec} denotes the azimuthal angle of the \textit{i}th particle (obtained from particle's momentum) in the event plane determination and $\omega_{i}$ is $p_{T}$ of the \textit{i}th particle. The reaction plane distribution should be isotropic. Due to the finite detector efficiency and acceptance, the detected particles are azimuthal anisotropic in the laboratory system which leads to an anisotropic distribution of reconstructed event plane distribution~\cite{Poskanzer:1998yz,Voloshin:2008dg}. The black line in Fig.~\ref{fig:ep}(a) presents the raw $\Psi_{1}$ distribution observed from MWDC. 

To correct the effect of anisotropic $\Psi_{1}$ distribution, the re-centering correction is applied~\cite{PhysRevC.56.3254}. The 1st-order flow vector $\boldsymbol{Q}_{1}$ is recalculated by subtracting the average divination between the distribution of $(Q_{x,1},Q_{y,1})$ and $(0,0)$, as described by
\begin{equation}
\label{eq:rc}
	\begin{gathered}
		\boldsymbol{q}_{\text{rec}} =\left( \begin{array}{c} {q}_{x,\text{rec}}\\{q}_{y,\text{rec}} \end{array} \right)=\left( \begin{array}{c}
		\left<\cos(\phi_{i})\right> \\ \left<\sin(\phi_{i})\right> \end{array} \right), \\
		\boldsymbol{Q}_{1}
		=\left( \begin{array}{c} \sum_{i}\omega_{i}(\cos(\phi_{i})-{q_{x,\text{rec}}})\\ \sum_{i}\omega_{i}(\sin(\phi_{i})-{q_{y,\text{rec}}})\end{array} \right), \\
	\end{gathered}
\end{equation}
where the $\boldsymbol{q}_{\text{rec}}$ is the average over all particles used in event plane determination from all events. 

The re-centered $\Psi_{1}$ distribution from MWDC is not perfectly flat as shown by the red line in Fig.~\ref{fig:ep}(a). The remaining anisotropic structure could be corrected by Shifting procedure~\cite{STAR:2021yiu}. For each event, a shift angle $\Delta\Psi_{1}$ could be calculated from the following equation:
\begin{equation}
\label{eq:shift}
\begin{gathered}
\Psi_{1}^{'} = \Delta{\Psi_{1}}+\Psi_{1},\\
\Delta{\Psi_{1}}=\sum_{i=1}^{n}\frac{2}{i}[-\langle\sin(i\Psi_{1})\rangle\cos(i\Psi_{1})+\langle\cos(i\Psi_{1})\rangle\sin(i\Psi_{1})],
\end{gathered}
\end{equation}
where $n$ is the maximum correction order and the brackets refer to the average over the events used in event plane reconstruction. $\Psi_{1}$ is the event plane angle after re-center correction and the $\Psi_{1}^{'}$ is the event plane angle after shift correction. As indicated by blue line in Fig.~\ref{fig:ep}(a), a flat event plane distribution reconstructed by MWDC is achieved after shift correction.

For hit-based ZDC, the $\phi_{i}$ used in Eq.~\ref{eq:qVec} denotes the azimuthal angle in the laboratory frame of the \textit{i}th particle hit on the ZDC and $\omega_{i}$ is the energy deposition $\Delta{E}$ of the \textit{i}th particle in the ZDC~\cite{Liu:2023xhc}. 
The magnetic field direction of CEE experiment is along the y-axis which perpendicular to the beam direction. The final state charged particles are deflected by the magnetic field, therefore, more likely to hit on one side of the ZDC, as shown in Fig.~\ref{fig:acpt}(c). This acceptance asymmetry on ZDC will bias the reconstructed event plane toward the $\pi$-azimuth as shown by the black line in Fig.~\ref{fig:ep}(b). To correct for this acceptance bias caused by the magnetic field, Ref.~\cite{Liu:2019nii} proposed a position weight correction to calibrate asymmetric acceptance as defined in Eq.~\ref{eq:poswgt}: 
\begin{equation}
\label{eq:poswgt}
\begin{gathered}
  w_{i} = \Delta{E}\times{P}, \\
  P = \left\{ \begin{array}{cc}
  n(-x,y,\Delta{E})/n(x,y,\Delta{E})&x<0, \\ 
  1  &x>0 ,\end{array}\right.  
\end{gathered}
\end{equation}
where weight $w_{i}$ is the deposited energy $\Delta{E}$ of the $i$th particle hit on the ZDC with an additional position weight factor based on two-dimensional X–Y hit distribution on ZDC. The position weight is calculated by the ratio of the number of hits on the right side of ZDC to that on the left side with the deposited energy $\Delta{E}$ as the weight~\cite{Liu:2019nii}. 
A shift correction is also needed after the position weight correction since the event plane distribution is not perfectly flat as shown by the red line in Fig.~\ref{fig:ep}(b). The shift angle is calculated from Eq.~\ref{eq:shift}, where $\Psi_{1}$ is the position-weight-corrected event plane angle, and $\Psi_{1}^{'}$ is the event plane angle after shift calibration. The blue line in Fig.~\ref{fig:ep}(b) presents the final ZDC event plane distribution with all corrections.

Due to the finite multiplicity of the final state particles, the reconstructed event plane deviates from the reaction plane, the average deviation could be estimated by the so-called event plane resolution. In this study, we focus on the simulation of proton $v_{1}$, thus, the $1^{\rm st}$-order event plane resolution is used in the simulation as defined in Eq.~\ref{eq:epRes}:
\begin{equation}
\label{eq:epRes}
    R_{1} = \left<\cos(\Psi_{1,\text{EP}}-\Psi_{\text{RP}})\right>.
\end{equation}

Since the $\Psi_{\text{RP}}$ is not directly measurable, the $1^{st}$-order event plane resolution from MWDC and ZDC are extracted with two-sub-event plane method~\cite{Poskanzer:1998yz,Voloshin:2008dg}. In this approach, each event used in the event plane resolution calculation is randomly divided into two sub-events with equal tracks (MWDC) or hits (ZDC). The event plane resolution of the two-sub-event could be calculated by 
\begin{equation}
\label{eq:epResSub}
R_{1,\text{sub}} = \sqrt{\left\langle\cos\left(\Psi_{1,\text{EP}}^{a}-\Psi_{1,\text{EP}}^{b}\right)\right\rangle}, 
\end{equation}
where $\Psi_{1,\text{EP}}^{a}$ and $\Psi_{1,\text{EP}}^{b}$ are the corrected event plane angle of of two sub-events. Then the full event plane resolution could be calculated from Eq.~\ref{eq:epResFull}:
\begin{equation}
\label{eq:epResFull}
R_{1} = \frac{\sqrt{\pi}}{2}\chi_{1} e^{\left(-\chi_{1}^{2}/2\right)}\left(I_{0}\left(\chi_{1}^{2} / 2\right)+I_{1}\left(\chi_{1}^{2}/2\right)\right),
\end{equation}
where $I_{0}$ and $I_{1}$ are the modified Bessel functions, and $\chi_{1} = v_{1}\sqrt{M}$ which is proportional to the square root of event multiplicity. Thus, the full event plane resolution could be obtained by $R_{1,\text{full}} = R_{1}(\sqrt{2}\chi_{1,\text{sub}})$~\cite{Liu:2023xhc,Poskanzer:1998yz}.

\begin{figure}[!htb]
\includegraphics[width=0.9\columnwidth]{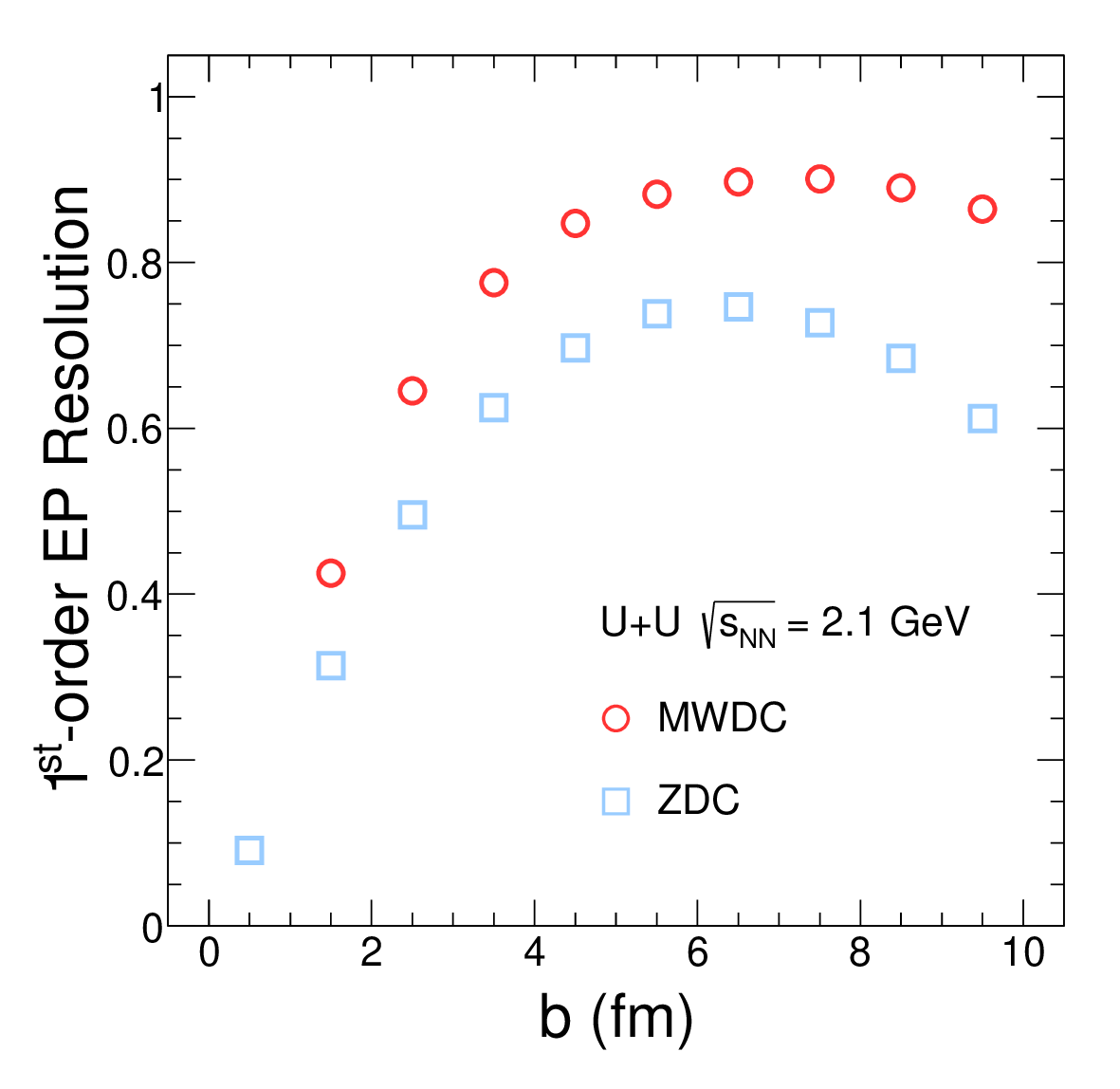}
\caption{1$^{\rm st}$-order event plane resolution as a function of impact parameter b from MWDC and ZDC.}
\label{fig:res}
\end{figure}

Figure~\ref{fig:res} presents $1^{\rm st}$-order event plane resolution as a function of impact parameter from MWDC and ZDC using two-sub-event method. In general, the event plane resolution from MWDC is higher than that from ZDC. Both detectors show a maximum resolution ($\sim$90\% for MWDC and $\sim$70\% for ZDC) in the mid-central collisions ($5 < b < 7$ fm). The absolute value of event plane resolution might be different with different input models, but the MWDC is in general having better $1^{\rm st}$-order event plane resolutions than the ZDC. One thing needs to mention here is the lack of event plane resolution of MWDC from $0 < b < 1$ fm. This is mainly due to the large non-flow effect which causes the negative correlation between two MWDC sub-events. This effect could be corrected from simulation in general, but beyond the scope of this paper.

\section{Directed Flow Simulation from TPC}
\label{sec:tpcFlow}

With the corrected event plane from MWDC (ZDC) and the corresponding event plane resolution, the directed flow of charged particles detected by TPC could be calculated with Eq.~\ref{eq:v1}:
\begin{equation}
    \label{eq:v1}
    v_{1} = \left<\cos(\phi_{i}^{\text{TPC}}-\Psi_{1,\text{EP}})\right>/R_{1} ,
\end{equation}
where $\phi_{i}^{\text{TPC}}$ is the azimuth angle of the particle of interest (POI) in TPC and the bracket means the average of all POIs within selected kinematic range in all events with the same event category. In this study, we select the events with impact parameter $5 < b < 6$ fm for illustration and POIs are protons within $0.2 < p_{T} < 0.7$ GeV/\textit{c} and $-0.5 < y < 0.5$ as indicated by the dashed box in Fig.~\ref{fig:acpt} (d).

\begin{figure}[!htb]
\includegraphics[width=0.8\columnwidth]{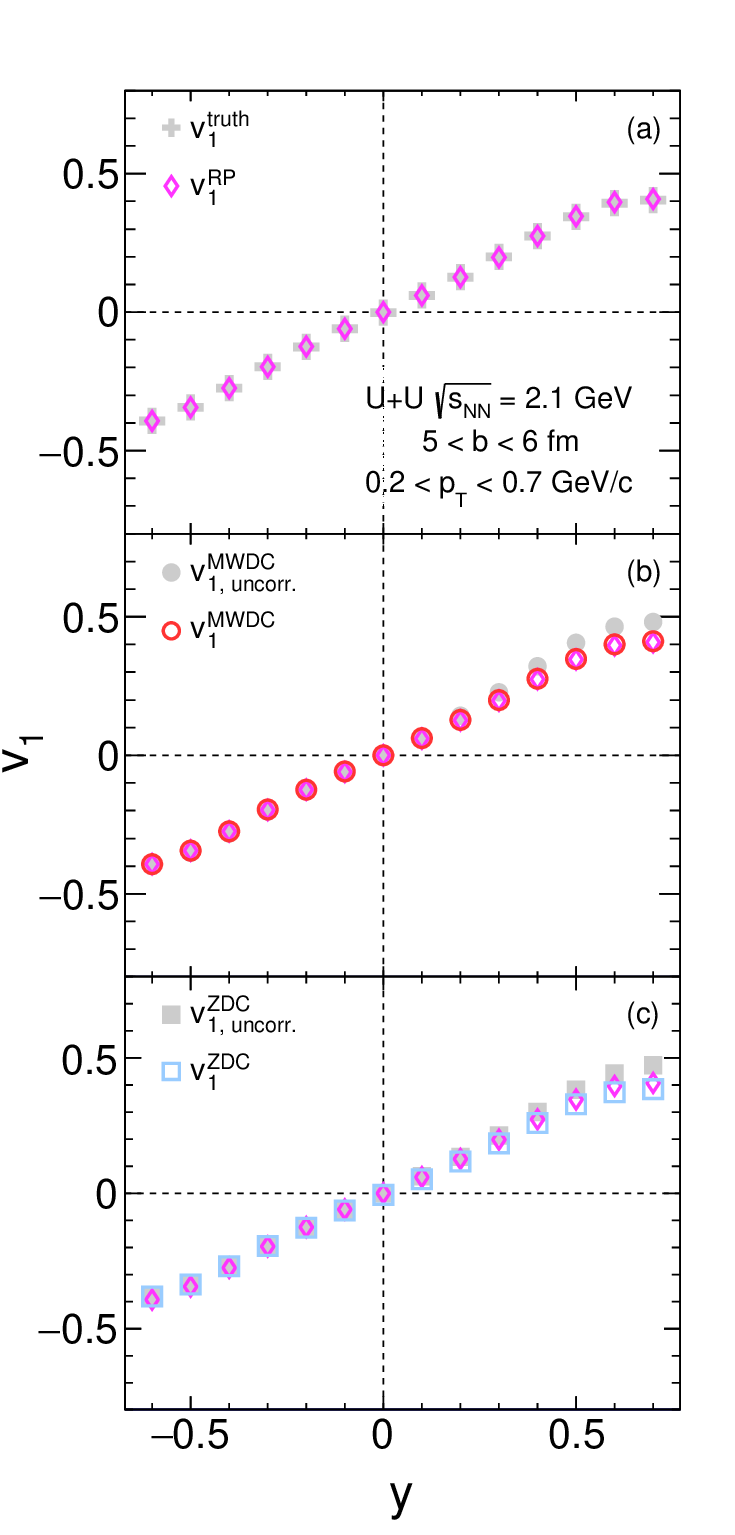}
\caption{Proton $v_{1}$ as a function of rapidity extracted from different detectors: (a) proton $v_{1}$ from JAM model (gray cross) and $v_{1}$ extracted from reaction plane from CFS package (magenta open diamonds); (b) un-corrected proton $v_{1}$ (gray filled circles) and corrected proton $v_{1}$ (red open circles, corrected for self-correlation and momentum conservation effect) from 1$^{\rm st}$-order MWDC event plane; (c) un-corrected proton $v_{1}$ (gray filled squares) and corrected proton $v_{1}$ (blue open squares, corrected for self-correlation) from 1$^{\rm st}$-order ZDC event plane.}
\label{fig:v1EP}
\end{figure}

Figure~\ref{fig:v1EP} presents the comparison of the proton $v_{1}$ signal calculated w.r.t. reaction plane and event plane reconstructed by different detectors. To validate the $v_{1}$ single measured by CEE detector, $v_{1}^{\text{RP}}$ (proton $v_{1}$ obtained from CFS package w.r.t. $\Psi_{\text{RP}}$) is compared with $v_{1}^{\text{truth}}$ (proton $v_{1}$ calculated from JAM model with the same kinetic acceptance as CFS). As shwon in Fig.~\ref{fig:v1EP}(a), $v_{1}^{\text{RP}}$ (magenta open diamond) is consistent with $v_{1}^{\text{truth}}$ (gray cross), which indicates the flow signal measured by CEE detector is reliable if $\Psi_{\text{RP}}$ is known. Fig.~\ref{fig:v1EP}(b) and (c) present the comparison between $v_{1}^{\text{MWDC}}$ (proton $v_{1}$ obtained from CFS package w.r.t. $\Psi_{\text{1,EP}}^{\text{MWDC}}$) and $v_{1}^{\text{ZDC}}$ (proton $v_{1}$ obtained from CFS package w.r.t. $\Psi_{\text{1,EP}}^{\text{ZDC}}$) to $v_{1}^{\text{RP}}$. After removing the self-correlation effect~\cite{Voloshin:2008dg,Poskanzer:1998yz} and momentum convservation effect~\cite{Borghini:2002mv}, $v_{1}^{\text{MWDC}}$ (red open cycles in Fig.~\ref{fig:v1EP}(b)) and $v_{1}^{\text{ZDC}}$ (blue open squres in Fig.~\ref{fig:v1EP}(c)) are consistent with $v_{1}^{\text{RP}}$ (magenta open diamonds), which means the \textit{standard event plane method} is valid and applicable for flow measurement in CEE experiment.

\begin{figure}[!htb]
\includegraphics[width=0.9\columnwidth]{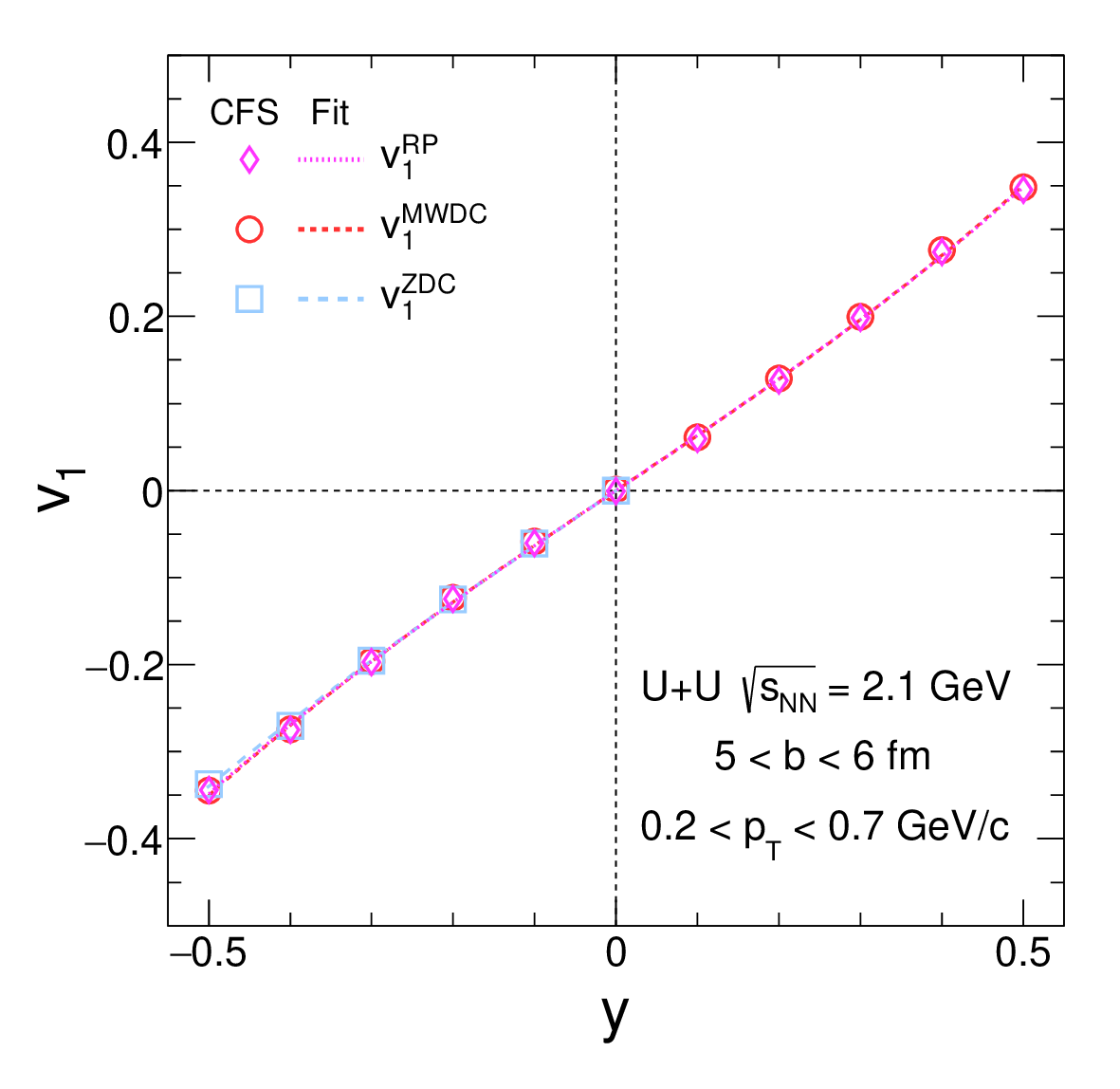}
\caption{Comparison of proton $v_{1}$ extracted from reaction plane (magenta open diamonds), 1$^{\rm st}$-order MWDC event plane (red open circles) and 1$^{\rm st}$-order ZDC event plane (blue open squares). The dashed lines are fits to proton $v_{1}$ extracted from corresponding planes. The fit function is $y = ax + bx^{3}$.}
\label{fig:v1Com}
\end{figure}

It is worth noting that the removal of self-correlation and momentum conservation effect is on a track-by-track basis~\cite{Voloshin:2008dg,Poskanzer:1998yz,Borghini:2002mv}, this is achievable by carefully matching reconstructed tracks from MWDC and TPC. But for ZDC, given the complicated magnetic filed and the position of ZDC, it is difficult to do a precise matching between ZDC hits and TPC/MWDC tracks. Thus, it would be complicated to remove such effects for $v_{1}^{\text{ZDC}}$. To accommodate such situation, we propose to use $\Psi_{1,\text{EP}}^{\text{ZDC}}$ only for the backward protons (or other charged particles) in TPC, i.e. $-0.5 < y < 0$, to avoid the self-correlation effect. The proposed measurement is shown in Fig.~\ref{fig:v1Com}. The dashed lines are fits to extract $v_{1}$ slope and the fit function is $y = ax + bx^{3}$. The $dv_{1}/dy$ is 0.631 $\pm$ 0.002 for $v_{1}^{\text{RP}}$, 0.629 $\pm$ 0.003 for $v_{1}^{\text{MWDC}}$ and 0.634 $\pm$ 0.012 for $v_{1}^{\text{ZDC}}$. The $v_{1}$ slopes extracted from $v_{1}^{\text{MWDC}}$ and $v_{1}^{\text{ZDC}}$ are consistent with that from $v_{1}^{\text{RP}}$ within 1$\sigma$, which means the proposed measurement region is reasonable for CEE experiment.

\begin{figure}[!htb]
\includegraphics[width=0.9\columnwidth]{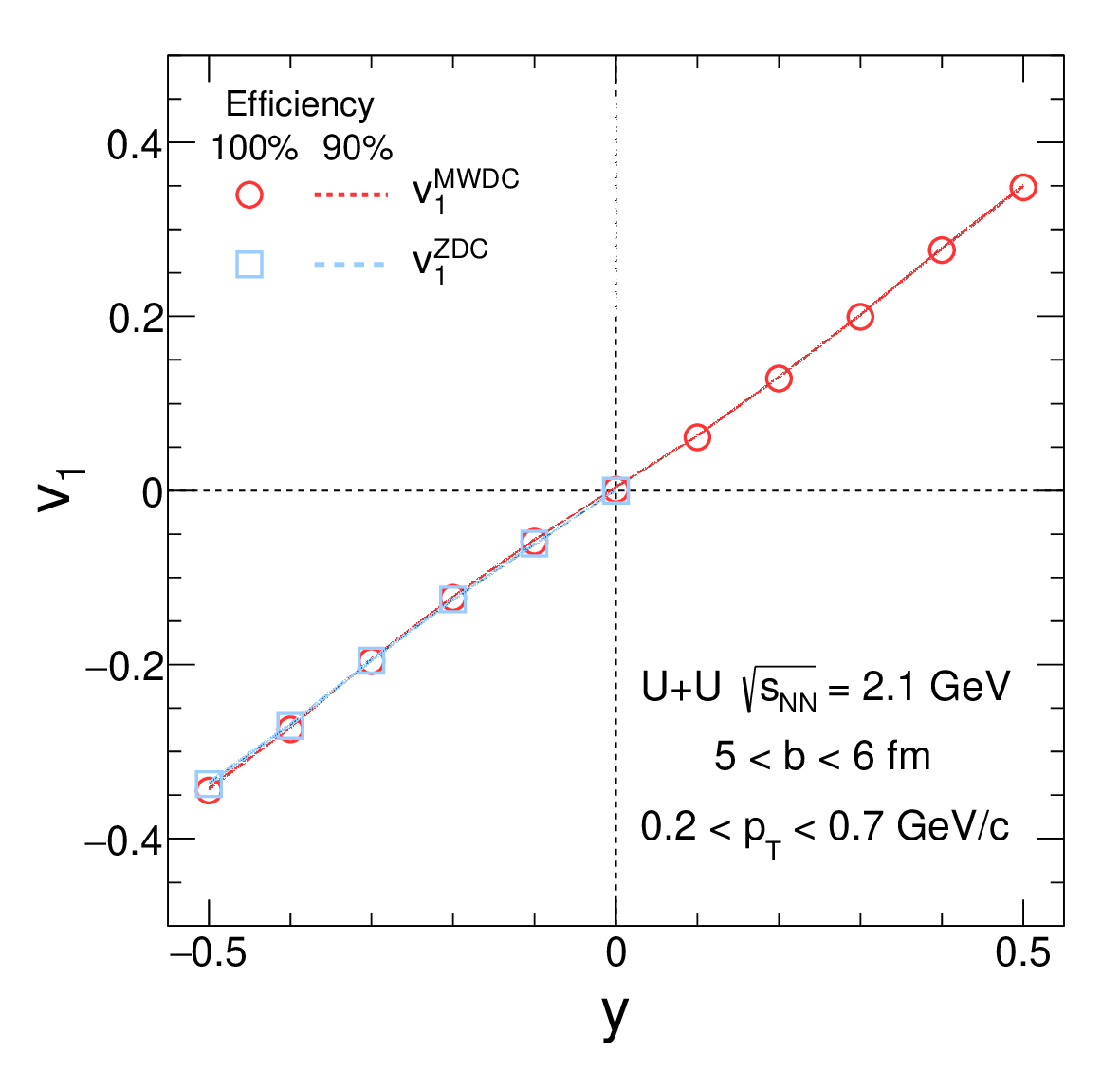}
\caption{Comparison of proton $v_{1}$ with different detector efficiency. Open symbols are proton $v_{1}$ extracted with 100$\%$ efficiency from MWDC (red open circles) and ZDC (blue open squares). Dashed lines are proton $v_{1}$ extracted with 90$\%$ efficiency from MWDC (red dashed line) and ZDC (blue dashed line).}
\label{fig:v1Eff}
\end{figure}

To further study the influence of detector effect on the flow measurement, different detector efficiency is introduced into CFS package. A universal 90\%  efficiency is applied to all relevant detectors for flow measurement such as TPC, MWDC and ZDC. The comparison between different efficiency is shown in Fig.~\ref{fig:v1Eff}, only 100\% (symbols) and 90\% (bands) efficiency are shown for simplicity. The consistency between $v_{1}$ obtained with 100\% and 90\% efficiency means the influence of detector efficiency is negligible for flow measurement in the CEE experiment. It needs to be pointed out that a realistic efficiency of each sub-detector should be applied for a complete study, but the general conclusion shouldn't be affected.

\section{Summary}
\label{sec:summary}
In summary, we presented the procedure of directed flow simulation using \textit{standard event plane method }in the CEE experiment. The simulation used JAM model (500 MeV/u $^{238}$U+$^{238}$U) as input and filtered by CFS package to provide CEE detector enviroments.
The charged particles detected by TPC are correlated with 1$^{\text{st}}$-order event plane reconstructed by MWDC and ZDC to extract the directed flow signal. The correction procedures of event plane and corresponding $v_{1}$ were also discussed in the paper.
The consistency among $v_{1}^{\text{MWDC}}$, $v_{1}^{\text{ZDC}}$, $v_{1}^{\text{RP}}$ and $v_{1}^{\text{truth}}$ indicated validity of \textit{standard event plane method} in the CEE experiment.
We also proposed the optimal kinematic region for $v_{1}$ measurement using 1$^{\rm st}$-order event plane reconstructed by MWDC and ZDC. 
The procedure and kinematic regions discussed in this study could be used as a guidance in the future CEE experiment.

\section{Acknowledgments}
We express our appreciation to Chuan Fu, Qichun Feng, Lei Huo, Qiang Hu, Li-Ke Liu, Hua Pei, Hao Qiu, Yaping Wang, Nu Xu, Jingbo Zhang and Yapeng Zhang for insightful discussions. 

\bibliographystyle{elsarticle-num} 
\bibliography{reference}

\end{document}